\documentclass[draft,numberedheadings]{aipproc}

\layoutstyle{8x11single}
\newcommand{\be}{\begin{equation}}
\newcommand{\ee}{\end{equation}}
\newcommand{\ba}{\begin{eqnarray}}
\newcommand{\ea}{\end{eqnarray}}

\date{}
\usepackage{amsfonts,amsbsy}
\begin{document}

\title[]{de Sitter Relativity and Quantum Physics}


\author{R. Aldrovandi}{
  address={Instituto de F\'{\i}sica Te\'orica,
Universidade Estadual Paulista \\
Rua Pamplona 145, 01405-900 S\~ao Paulo, Brazil}}

\author{J. P. Beltr\'an Almeida}{
  address={Instituto de F\'{\i}sica Te\'orica,
Universidade Estadual Paulista \\
Rua Pamplona 145, 01405-900 S\~ao Paulo, Brazil}}

\author{C. S. O. Mayor}{
  address={Instituto de F\'{\i}sica Te\'orica,
Universidade Estadual Paulista \\
Rua Pamplona 145, 01405-900 S\~ao Paulo, Brazil}}

\author{J. G. Pereira}{
  address={Instituto de F\'{\i}sica Te\'orica,
Universidade Estadual Paulista \\
Rua Pamplona 145, 01405-900 S\~ao Paulo, Brazil}}

\begin{abstract}
In the presence of a cosmological constant, interpreted as a purely geometric entity, absence of matter is represented by a de Sitter spacetime. As a consequence, ordinary Poincar\'e special relativity is no longer valid and must be replaced by a de Sitter special relativity. By considering the kinematics of a spinless particle in a de Sitter spacetime, we study the geodesics of this spacetime, the ensuing definitions of canonical momenta, and explore possible implications for quantum mechanics.
\end{abstract}
\maketitle

\section{Introduction}

When the cosmological constant $\Lambda$ vanishes, absence of gravitation is represented by Minkowski spacetime, a solution of the sourceless Einstein's equation. Its isometry transformations are those of the Poincar\'e group, which is the group governing the kinematics of special relativity. For a non-vanishing $\Lambda$, however, Minkowski is no longer a solution of the corresponding $\Lambda$--modified Einstein equation and becomes, in this sense, physically meaningless. In this case, if we interpret $\Lambda$ as a purely geometric entity, absence of gravitation turns out to be represented by the de Sitter spacetime. Now, the group governing the kinematics in a de Sitter spacetime is not the Poincar\'e, but the de Sitter group. This means that, in the presence of a non-vanishing $\Lambda$, ordinary Poincar\'e special relativity will no longer be valid, and must be replaced by a de Sitter special relativity~\cite{dssr}.\footnote{Similar ideas have been explored in Ref.~\cite{guoatall}.}

An important point of this theory is that it retains the quotient character of spacetime and, consequently, a notion of homogeneity. As in ordinary special relativity, whose underlying Minkowski spacetime $M$ is the quotient space of the Poincar\'e by the Lorentz groups, the underlying spacetime of the de Sitter relativity will be the quotient space of the de Sitter and the Lorentz groups. Similarly to ordinary special relativity, therefore, in a de Sitter special relativity the Lorentz subgroup remains responsible for both the isotropy of space (rotation group) and the equivalence of inertial frames (boosts)~\cite{lorentz}. The four additional transformations, given by a combination of translations and proper conformal transformations, define the homogeneity of spacetime.

A space is said to be transitive under a set of transformations --- or homogeneous under them --- when any two points of it can be attained from each other by a transformation belonging to the set. For example, Minkowski spacetime is transitive under spacetime translations. However, the de Sitter spacetime is transitive, not under translations, but under a combination of 
translations and proper conformal transformations, with the relative importance of 
these contributions being determined by the value of the cosmological constant. 
An immediate consequence of this property is that the ordinary notions of energy and momentum will change~\cite{aap}. In fact, the conserved momentum, for example, will now be obtained from the invariance of the physical system, not under translations, but under a combination of translations and proper conformal transformations. The conserved momentum, therefore, will be a combination of ordinary and proper conformal momenta~\cite{confor}.

Due to the smallness of the observed cosmological constant, the difference between ordinary and de Sitter relativities will be very small. However, there are situations where this difference could become significant. For example, according to our current theories on fields and particles, the phase transitions associated to the spontaneously broken symmetries can be considered the primary source for a non-vanishing $\Lambda$~\cite{carroll}. According to this view, a high energy experiment modifies the {\it local structure of space-time for a short period of time}, in such a way that the immediate neighborhood of a high energy collision departs from Minkowski and becomes a de Sitter --- or anti-de Sitter --- spacetime~\cite{mansouri}. There would then exist a connection between the energy scale of the experiment and the local value of $\Lambda$. The higher the energy, the larger the local value of $\Lambda$, and consequently the large the importance of the conformal symmetry. This is consistent with the idea that, at high energies the masses become negligible, thence the importance of the conformal symmetry. For a high enough energy, the local kinematics would be ruled preponderantly by a de Sitter special relativity.

The basic purpose of these notes is to study the kinematics subjacent to a de Sitter special relativity. In particular, we are going to study the equation of motion of spinless particles, which defines the geodesics of the de Sitter spacetime. Then, through an usual analysis of the action functional variation, we obtain the canonical momentum, and explore the consequences of this definition for quantum mechanics. We begin by introducing first the basic notions associated to the de Sitter spacetimes and groups.

\section{de Sitter spacetimes and groups}

Spacetimes with constant scalar curvature $R$ are maximally symmetric: they can 
lodge the highest possible number of Killing vectors. Given a metric signature, 
this spacetime is unique~\cite{weinberg} for each value of $R$. Minkowski 
spacetime $M$, with $R = 0$, is the simplest one. Its group of motions is the 
Poin\-ca\-r\'e group ${\mathcal P} = {\mathcal L} \oslash {\mathcal T}$, the semi-
direct product of the Lorentz ${\mathcal L} = SO(3,1)$ and the translation group 
${\mathcal T}$. The latter acts transitively on $M$ and its group manifold can be 
identified with $M$. Indeed,  Minkowski  spacetime is a homogeneous space under 
${\mathcal P}$:
\[
M = {\mathcal P}/{\mathcal L}.
\]

Amongst curved spacetimes, the de Sitter and anti-de Sitter spaces are the only 
possibilities. One of them has negative, and the other has positive 
scalar curvature. They are hyper-surfaces in the ``host'' pseudo-Euclidean spaces 
${\bf E}^{4,1}$ and ${\bf E}^{3,2}$, inclusions whose points in Cartesian 
coordinates $(\chi^A) = (\chi^0,
\chi^1, \chi^2, \chi^3, \chi^{4})$ satisfy respectively~\cite{he}
\[
\eta_{AB} \chi^A \chi^B \equiv (\chi^0)^2 - (\chi^1)^2 -
(\chi^2)^2 - (\chi^3)^2 - (\chi^{4})^2 = -\, l^2
\]
and
\[
\eta_{AB} \chi^A \chi^B \equiv (\chi^0)^2 - (\chi^1)^2 -
(\chi^2)^2 - (\chi^3)^2 + (\chi^{4})^2 = l^2,
\]
where $l$ is the so-called de Sitter length-parameter. Using the Latin alphabet 
($a, b, c \dots = 0,1,2,3$) to denote the four-dimensional algebra and tangent 
space indices, whose metric tensor is $\eta_{a b} = $ diag $(1$, $-1$, $-1$, $-
1)$, and writing $\eta_{44} = {\sf s}$, the above conditions can be put together as
\be
\eta_{a b} \, \chi^{a} \chi^{b} + {\sf s} \, (\chi^{4})^2 
= {\sf s} \, l^2.
\label{dspace1}
\ee
For ${\sf s} = - 1$, we have the de Sitter space $dS(4,1)$, whose metric is 
induced from the pseudo-Euclidean metric $\eta_{AB}$ = $(+1,-1,-1,-1,-1)$. It has 
the pseudo-orthogonal group $SO(4,1)$ as group of motions. Sign ${\sf s} = + 1$ 
corresponds to anti-de Sitter space, denoted by $dS(3,2)$. It comes from 
$\eta_{AB}$ = $(+1,-1,-1,-1,+1)$, and has $SO(3,2)$ as its group of motions. Both 
spaces are homogeneous~\cite{livro}:
\[
dS(4,1) = SO(4,1)/ {\mathcal L} \quad {\rm and} \quad dS(3,2) = 
SO(3,2)/ {\mathcal L}.
\]
Furthermore, they are solutions of the sourceless Einstein's equation, provided 
the cosmological constant $\Lambda$ and the de Sitter length-parameter $l$ are 
related by
\be
\Lambda = -\, \frac{3 {\sf s}}{l^2}.
\label{lambdaR}
\ee

We consider from now on the $dS$ spacetime, for which $\Lambda > 0$. In this case, the four-dimensional stereographic coordinates $\{x^a\}$ --- which are obtained through a stereographic
projection from the de Sitter hyper--surface into a target Minkowski spacetime --- is given by \cite{gursey}
\be
\chi^{a} = \Omega(x) \, x^a \quad \mbox{and} \quad
\chi^4 = -\, l \, \Omega(x) \left(1 + \frac{\sigma^2}{4 l^2} \right),
\label{xi4}
\ee 
where
\be
\Omega(x) = \left(1 - \frac{\sigma^2}{4 l^2}\right)^{-1},
\label{n}
\ee
with $\sigma^2 = \eta_{a b} \, x^a x^b$ a Lorentz invariant quadratic interval. The inverse relations are
\be
x^a = \Omega^{-1}(\chi) \, \chi^a \quad \mbox{and} \quad
\frac{\sigma^2}{4 l^2} = \frac{(\chi^4/l) + 1}{(\chi^4/l) - 1},
\label{inverse}
\ee 
where now
\be
\Omega(\chi) = \frac{1}{2} \left( 1 - \frac{\chi^4}{l} \right).
\ee
In terms of the stereographic coordinates, the five--dimensional line element
\be
d\tau^2 = \eta_{AB} \, d \chi^A d \chi^B
\label{dtau1}
\ee
reduces to the de Sitter line element
\be
d\tau^2 = g_{ab} \, dx^a dx^b,
\label{dsle}
\ee
with
\be
g_{ab} = \Omega^2 \, \eta_{ab}
\label{dSmetric}
\ee
the metric of de Sitter spacetime.

The generators of infinitesimal de Sitter transformations are
\be
{\mathcal L}_{A B} = \eta_{AC} \, \chi^C \, \frac{\partial}{\partial \chi^B} -
\eta_{BC} \, \chi^C \, \frac{\partial}{\partial \chi^A}.
\label{dsgene}
\ee
In terms of the stereographic coordinates $\{x^a\}$, these generators are written 
as
\be
{\mathcal L}_{ab} =
\eta_{ac} \, x^c \, {\mathcal P}_b - \eta_{bc} \, x^c \, {\mathcal P}_a
\label{cp0}
\ee
and
\be
{\mathcal L}_{a4} = l {\mathcal P}_a - ({4 l})^{-1} {\mathcal K}_a,
\label{dstra}
\ee
where
\be
{\mathcal P}_a = {\partial}/{\partial x^a}
\ee
are the translation generators (with dimension of {\it length}$^{-1}$), and
\be
{\mathcal K}_a = \left(2 \eta_{ab} \, x^b x^c - \sigma^2 \delta_{a}{}^{c} \right)
{\partial}/{\partial x^c}
\ee
are the generators of {\it proper} conformal transformations (with dimension of {\it length}). Generators ${\mathcal L}_{a b}$ refer to the Lorentz subgroup of de Sitter, whereas ${\mathcal L}_{a4}$ define transitivity on the corresponding de Sitter space. For this reason, they are usually called the de Sitter ``translation'' generators. As implied by Eq.~(\ref{dstra}), the de Sitter spacetime is seen to be transitive under a combination of translations and proper conformal transformations. The relative importance of each one of these transformations is determined by the value of the length parameter $l$, that is, by the value of the cosmological constant. 

The group contraction procedure requires that, before each limit is taken, the generators be modified through an appropriate insertion of parameters. These alterations are frequently guided by dimensional considerations and are different for different limits \cite{inonu2}. For this reason, the limits of small and large cosmological constant must be considered separately. In particular, for a small cosmological constant, it is convenient to rewrite the de Sitter generators in the forms
\be
{\mathcal L}_{ab} =
\eta_{ac} \, x^c \, {\mathcal P}_b - \eta_{bc} \, x^c \, {\mathcal P}_a
\label{dslore}
\ee
and
\be
{\Pi}_a \equiv \frac{{\mathcal L}_{a4}}{l} =
{\mathcal P}_a - \frac{1}{4 l^2}\,  {\mathcal K}_a.
\label{l0}
\ee
For $l \to \infty$, the generators $\Pi_a$ reduce to ordinary translations, and 
the de Sitter group contracts to the Poincar\'e group ${\mathcal P} = {\mathcal L} 
\oslash {\mathcal T}$. Concomitant with the algebra and group deformations, the de 
Sitter space $dS(4,1)$ reduces to the Minkowski spacetime $M = {\mathcal 
P}/{\mathcal L}$, which is transitive under ordinary translations only.

\section{Kinematics in Minkowski Spacetime Revisited}

The action functional describing a free particle of mass $m$ moving in a Minkowski spacetime is
\be
S = - m c \int_a^b ds,
\label{aM1}
\ee
where
\be
ds = (\eta_{ab} \, dx^a dx^b)^{1/2}
\ee
is the Lorentz invariant interval. Now, the kinematic group of Minkowski is the Poincar\'e group ${\mathcal P} = {\mathcal L} \oslash {\mathcal T}$, the semi-direct product of the Lorentz ${\mathcal L}$ and the translation group ${\mathcal T}$. The first Casimir invariant of the Poincar\'e group, on the other hand, is
\be
{\mathcal C}_P = \eta_{ab} \, p^a p^b = m^2 c^2
\label{cons1}
\ee
where $p^a = m c u^a$ is the particle four-momentum, with $u^a = dx^a/ds$ the four--velocity. Considering that the action $S$ and the Lagrangian $L$ are related by
\be
S = \frac{1}{c} \int_a^b {L} \; ds,
\ee
the corresponding Lagrangian can then be written in the form
\be
{L} = - \, c \, (\eta_{ab} \, p^a p^b)^{1/2} = - \, c \, \sqrt{{\mathcal C}_P}.
\label{L1}
\ee
The identity $\eta_{ab} \, p^a p^b = m^2 \, c^2$ is a weak constraint in the sense that it can be used only after the variational calculus is performed. The resulting equation of motion is
\be
\frac{d p^a}{ds} = 0.
\ee
The equation of motion, therefore, coincides with the conservation of the particle four-momentum, which follows from the invariance of the system under spacetime translation. Its solution determines the geodesics of the Minkowski spacetime. The invariance of the system under Lorentz transformations yields the conservation of the particle angular momentum $\lambda^{ab} = x^a p^b - x^b p^a$, that is,
\be
\frac{d \lambda^{ab}}{ds} = 0.
\ee

\section{Kinematics in de Sitter Spacetime}

\subsection{Casimir Invariant}

For a spinless particle of mass $m$, the first Casimir invariant of the de Sitter group is given by~\cite{gursey}
\be
{\mathcal C}_{dS} = - \frac{1}{2 l^2} \, \eta_{AC} \, \eta_{BD} \, \lambda^{AB} \, \lambda^{CD},
\label{dsCas}
\ee
where
\be
\lambda^{A B} = m \, c \left(\chi^A \; \frac{d \chi^B}{d \tau} - \chi^B \; \frac{d 
\chi^A}{d \tau} \right)
\ee
is the conserved five-dimensional angular momentum. In terms of the stereographic coordinates $\{x^a\}$, the Casimir invariant (\ref{dsCas}) assumes the form
\be
{\mathcal C}_{dS} = \eta_{ab} \, \pi^a \, \pi^b -
\frac{1}{2 l^2} \, \eta_{ac} \, \eta_{bd} \, \lambda^{ab} \, \lambda^{cd},
\label{cas1}
\ee
where\footnote{Analogously to the generators, we use a parameterization appropriate for a small cosmological constant.}
\be
\pi^a \equiv \frac{\lambda^{a4}}{l} = \Omega^2 \Big( {p}^a - \frac{{k}^a}{4 l^2} \Big), \qquad
\pi_a \equiv \frac{\lambda_a{}^{4}}{l} = {p}_a - \frac{{k}_a}{4 l^2}
\label{dstra8}
\ee
represents the de Sitter momentum, with
\be
p^a = m \, c \, \frac{dx^a}{d\tau} \qquad \mbox{and} \qquad
k^a \equiv \bar{\delta}^a{}_b \; p^b = (2 \eta_{cb} \, x^c \, x^a -
\sigma^2 \, \delta^a{}_b) \,  {p}^b,
\label{pek}
\ee
respectively, the linear and the conformal momentum,\footnote{Similarly to the identification $p^a = T^{a 0}$, with $T^{ab}$ the energy-momentum tensor, the conformal momentum $k^a$ is defined by $k^a = K^{a 0}$, with $K^{ab}$ the conformal current \cite{coleman}.} and
\be
\lambda^{ab} = \Omega^2 \left( x^a \, p^b - x^b \, p^a \right), \qquad
\lambda_{ab} = \eta_{ac} \, x^c \, p_b - \eta_{bc} \, x^c \, p_a
\label{cc0}
\ee
represents the orbital angular momentum. In the above expression,
\be
\bar{\delta}^a{}_b = 2 \eta_{bc} \, x^c \, x^a -
\sigma^2 \, \delta^a{}_b
\label{khat}
\ee
is a kind of conformal Kroenecker delta. Since $\lambda^{A B}$ is conserved, we have also
\be
\frac{d \lambda^{ab}}{d \tau} = 0 \qquad \mbox{and} \qquad
\frac{d \pi^a}{d \tau} = 0.
\label{cls} 
\ee
We remark that $\lambda^{ab}$ is the Noether conserved momentum related to the invariance of the system under the transformations generated by ${\mathcal L}_{ab}$, whereas $\pi^a$ is the Noether conserved momentum related to the invariance of the system under the transformations generated by $\Pi_{a}$.\footnote{The same conservation laws, but in different coordinates, were studied in ref.~\cite{ugo}.}

\subsection{Equations of Motion}

Relying on the Minkowski case, the Lagrangian of a spinless particle of mass $m$ in de Sitter spacetime can be assumed to be given by  $- c \sqrt{{\mathcal C}_{dS}}$. In the five--dimensional spacetime, however, it is necessary to add a constraint restricting the movement to the de Sitter hyperboloid. In this case, therefore, the Lagrangian turns out to be
\be
L = - c \left[ \big( {\mathcal C}_{dS} \big)^{1/2} + \;
\beta \left(\eta_{AB} \, \chi^A \chi^B + l^2 \right) \right],
\ee
where $\beta$ is a Lagrange multiplier. Using Eq.~(\ref{dsCas}), the corresponding action is written as
\be
{S} = - \int_a^b \left[ \Big( - \frac{1}{2 l^2} \, \eta_{AC} \, \eta_{BD} \, \lambda^{AB} \, \lambda^{CD} \Big)^{1/2} +
\beta \left(\eta_{AB} \, \chi^A \chi^B + l^2 \right) \right] d\tau,
\label{action1}
\ee
with $d\tau$ the de Sitter invariant interval (\ref{dtau1}). Performing a functional variation, and neglecting the surface term coming from an integration by parts, the invariance of the action yields the equation of motion
\be
\frac{d^2 \chi^A}{d \tau^2} + \Big(\frac{1}{l^2} - 2 \, \beta \Big) \, \chi^A = 0.
\ee
Using the constraints
\be
\eta_{AB} \; \chi^A \chi^B = - \, l^2 \qquad \mbox{and} \qquad 
\eta_{AB} \; \frac{d \chi^A}{d \tau} \frac{d \chi^B}{d \tau} = 1,
\ee
the value of the Lagrange multiplier is found to be
\be
\beta = \frac{1}{l^2},
\label{lmult}
\ee
and the equation of motion reduces to to
\be
\frac{d^2 \chi^A}{d \tau^2} - \frac{\chi^A}{l^2} = 0.
\ee
In terms of the stereographic coordinates, it becomes
\be
\frac{d }{d \tau} \left(p^a + \frac{1}{4 l^2} \, k^a \right) +
\frac{x^c \, u_c}{l^2 \, \Omega} \, \left(p^a + \frac{1}{4 l^2} \, k^a \right) -
\frac{m \, c}{l^2 \, \Omega} \; x^a = 0.
\label{masterEq}
\ee
The corresponding equation for the covariant components of the momentum is
\be
\frac{d }{d \tau} \left(p_a + \frac{1}{4 l^2} \, k_a \right) -
\frac{m \, c \, \Omega}{l^2} \; \eta_{ab} \, x^b = 0.
\label{comasterEq}
\ee
Of course, due to the universality of gravitation, this equation is independent of the mass when written in terms of the four velocity. In fact, it is the same as
\be
\frac{d }{d \tau} \left(u_a + \frac{1}{4 l^2} \, \bar{\delta}_a{}^c \, u_c \right) -
\frac{\Omega}{l^2} \; \eta_{ab} \, x^b = 0,
\label{comasterEq2}
\ee
with $\bar{\delta}_a{}^c$ given by Eq.~(\ref{khat}). This is the equation of motion of a spinless particle of mass $m$ in a de Sitter spacetime. Its solutions determine the geodesics of this spacetime.

Using the second of the conservation laws (\ref{cls}), it is possible to obtain separate evolution equations for $p^a$ and $k^a$. For example, the equation of motion for the linear momentum $p^a$ is found to be
\be
\frac{dp_a}{d \tau} - \frac{m \, c \, \Omega}{2 \, l^2} \; \eta_{ab} \, x^b = 0,
\label{dsem1}
\ee
or equivalently,
\be
\frac{d p^a}{d \tau} + \frac{x^c \, u_c}{l^2 \, \Omega} \; p^a -
\frac{m \, c}{2 l^2 \, \Omega} \; x^a = 0.
\label{dsem1bis}
\ee
This equation is nothing but the geodesic equation
\be
\frac{d p^a}{d \tau} + \Gamma^a{}_{bc} \, p^b \, u^c = 0,
\ee
with $\Gamma^a{}_{bc}$ the Levi--Civita connection of the de Sitter metric (\ref{dSmetric}). On the other hand, the equation of motion for the conformal momentum $k^a$ assumes the form
\be
\frac{d k^a}{d \tau} + \frac{x^c \, u_c}{l^2 \, \Omega} \; k^a -
\frac{2 \, m \, c}{\Omega} \; x^a = 0.
\label{dsem2}
\ee
Differently from the ordinary momentum $p^a$, which is conserved with a covariant derivative, the conformal momentum $k^a$ is not covariantly conserved. In fact, it is found to satisfy
\be
\frac{d k^a}{d \tau} + \Gamma^a{}_{bc} \, k^b \, u^c =
\frac{2 m c}{\Omega} \left[1 - \frac{1}{4 l^2} \left(\frac{2}{\Omega^2} \, u_b \, u_c \, x^b \, x^c - \sigma^2 \right) \right] \, x^a.
\ee
Put together, however, these two momenta yield the truly conserved total momentum $\pi^a$.

\section{The Scalar Field}

Let us consider now a de Sitter scalar field $\phi$, that is, a field invariant under de Sitter transformations. We begin by considering the Casimir invariant (\ref{cas1}), but now written as an operator, that is,
\begin{equation}
{\mathcal C}_{dS} = \eta^{ab} \, \Pi_a \, \Pi_b -
\frac{1}{2 l^2} \, \eta^{ac} \, \eta^{bd} \, {\mathcal L}_{ab} \, {\mathcal L}_{cd},
\label{cas2}
\end{equation}
with $\Pi_a$ and ${\mathcal L}_{ab}$ the de Sitter generators, given respectively by Eqs.~(\ref{dslore}) and (\ref{l0}). From the theory of group representations, one finds~\cite{dix}
\begin{equation}
{\mathcal C}_{dS} = m^2 \, c^2 + \frac{1}{l^2} \, \left[s(s+1)-2 \right],
\label{rforcs}
\end{equation}
with $m$ the mass and $s$ the spin of the field ($\hbar = 1$). For a scalar field, $s = 0$, and we get\footnote{Sometimes $\alpha(m)$ is called ``de Sitter mass''. However, we prefer not to use this terminology because, strictly speaking, it does not represent a mass~\cite{massnomass}. The only mass present is the physical mass $m$.}
\begin{equation}
{\mathcal C}_{dS} =
\left( m^2 - \frac{2}{l^2 \, c^2} \right) c^2 \equiv \alpha^2(m) \, c^2.
\label{dSmass}
\end{equation}

Now, for a scalar field, we can identify
\be
{\mathcal C}_{dS} = - \, \Box,
\ee
where
\be
\Box =  - \, g_{ab} \, \nabla^a \nabla^b
\ee
is the d'Alembertian, with $\nabla^a$ the covariant derivative in the Levi--Civita connection of the de Sitter metric (\ref{dSmetric}). Accordingly, the field equation for $\phi$ is found to be
\begin{equation}
\Box \phi + m^2 c^2 \, \phi - \frac{R}{6} \, \phi = 0,
\label{cskg}
\end{equation}
where $R = 12/l^2$ is the scalar curvature of the de Sitter spacetime. For a massless field, it reduces to the conformal invariant field equation for $\phi$. This shows in a simple manner that the cosmological constant naturally introduces the conformal symmetry in any physical problem. This symmetry is enforced by the second term of the Casimir operator (\ref{rforcs}). In fact, it is responsible for the curvature term in the Klein--Gordon equation (\ref{cskg}), which is essential for the conformal invariance of the equation. Observe that, for the case of the electromagnetic field ($s=1$), which is naturally conformal invariant in the absence of sources, this term does not contribute to the corresponding field equation.

Equation (\ref{cskg}), with zero mass and no self--interaction, has already been obtained in the study of conformally invariant equations for massless particles~\cite{penrose,deser}. It also appeared in connection with the so called improved energy--momentum tensor for the scalar field~\cite{ccj}. Here, it has been obtained simply by considering that, instead of a Lorentz scalar, the field is in a singlet representation of the de Sitter group.

\section{Quantum Physics}

Let us consider again the action (\ref{action1}). Its total variation is given by
\be
\delta S = m \, c \int_a^b \left[ \frac{d^2 \chi_A}{d \tau^2} -
\frac{\chi_A}{l^2} \right] \delta \chi^A \, d\tau -
\int_a^b \left[m \, c \, \frac{d \chi_A}{d \tau} \; \delta \chi^A \, \right],
\ee
where we have already used Eq.~(\ref{lmult}). If we admit now only possible trajectories, the first term in the variation vanishes identically. Then, the second term, with the upper limit considered as variable, gives the differential of the action as a function of the coordinates:
\be
\delta S = - \; m \, c \; \frac{d \chi_A}{d \tau} \; \delta \chi^A.
\ee
In terms of the stereographic coordinates, it reduces to
\be
\delta S = - \; p_b \, \delta x^b.
\ee
Consequently, we can write
\be
\frac{\delta S}{\delta x^b} = - \; p_b,
\ee
where
\be
p_b = m \, c \, u_b
\label{canomomen}
\ee
is the canonical momentum conjugate to the coordinate $x^b$.

We define now the quantum mechanical operator
\be
\hat{p}_b = - i \, \hbar \, \partial_b.
\ee
As is well known, it satisfies the commutation relation
\be
[\hat{x}^a, \hat{p}_b ] = - \; i \, \hbar \, \delta^a{}_b.
\label{ocr}
\ee
Analogously to the momentum, we define the operator
\be
\hat{k}_b = \bar{\delta}_b{}^c \, \hat{p}_c,
\ee
with $\bar{\delta}_b{}^c$ given by Eq.~(\ref{khat}). Using the fundamental commutation relation (\ref{ocr}), it is easy to see that
\be
[\,\hat{x}^a, \hat{k}_b ] = - \; i \, \hbar \, \bar{\delta}^a{}_b.
\label{ccr}
\ee
Now, as is well known, under a spacetime inversion
\be
x^a \rightarrow y^a = - \, \frac{x^a}{\sigma^2},
\ee
the translation generators are transformed into the proper conformal generators, and {\it vice--versa}~\cite{coleman}. Using this property, we obtain
\be
[\hat{y}^a, \hat{k}_b ] = - \; i \, \hbar \, \delta^a{}_b.
\label{ccrbis}
\ee
This means that the conformal momentum $k_b$ is the canonical momentum conjugate to the coordinate $y^a$.

It is important to observe that the total momentum of the particle is given by $\pi^a = p^a - (1/4l^2) \, k^a$, with $p^a$ representing the part related to translations, and $k^a$ representing the part related to the proper conformal transformations. This follows from the fact that the de Sitter spacetime is transitive under a combination of translations and proper conformal transformations. The total energy of the particle, therefore, will be given by the time component of $\pi^a$, that is,
\be
\frac{E}{c} \equiv \pi^0 = p^0 - \frac{1}{4 l^2} \; k^0.
\ee
In other words,
\be
E = E_p - \frac{1}{4 l^2} \; E_k,
\ee
with $E_p = p^0$ and $E_k = k^0$. Accordingly, the total momentum operator is
\be
\hat{\pi}_b = \hat{p}_b - \frac{1}{4 l^2} \, \hat{k}_b,
\ee
which satisfies the commutation relation
\be
[\,\hat{x}^a, \hat{\pi}_b ] = - \; i \, \hbar \Big( \delta^a{}_b -
\frac{1}{4 l^2} \, \bar{\delta}^a{}_b \Big).
\ee
This is the quantization rule in de Sitter special relativity. Of course, since $\pi_b$ is not the conjugate momentum to the coordinate $x^a$, the right--hand side is not a Kroenecker delta.

The commutation rules, as is well known, are used to construct the uncertainty relations of quantum mechanics. For example, the commutation relation (\ref{ocr}) implies that
\be
\Delta x^a \; \Delta {p}_b \geq \frac{1}{2}\left| \left< [\, \hat{x}^a, \hat{p}_b]\right> \right| =
\frac{\hbar}{2} \; \delta^{a}{}_{b}.
\label{uncertp}
\ee
Analogously, the commutation relation (\ref{ccr}) implies that
\be
\Delta x^a \; \Delta {k}_b \geq \frac{1}{2}\left| \left< [\, \hat{x}^a, \hat{k}_b]\right> \right| =
\frac{\hbar}{2} \; \bar{\delta}^{a}{}_{b}.
\label{uncertk}
\ee
The uncertainty relation for the total momentum, consequently, is
\be
\Delta x^a \; \Delta {\pi}_b \geq
\frac{1}{2} \left| \left< [\, \hat{x}^a, \hat{\pi}_b]\right> \right| =
\frac{\hbar}{2} \, \Big( \delta^{a}{}_{b} - \frac{1}{4 l^2} \, \bar{\delta}^{a}{}_{b} \Big).
\label{moduncert}
\ee
For small values of the cosmological term $\Lambda$, the corrections to ordinary quantum mechanics will be very small. However, for large values of $\Lambda$, the corrections coming from the conformal momentum will become important, giving rise to a new quantum mechanics.

\section{Final Remarks}

A non-vanishing cosmological term $\Lambda$ introduces the conformal generators in the definition of spacetime transitivity. As a consequence, the conformal transformations will naturally be incorporated in the kinematics of spacetime, and the corresponding conformal current will appear as part of the Noether conserved current. Of course, for a small cosmological term, the conformal modifications become negligible and ordinary physics remains valid. For large values of $\Lambda$, however, the conformal contributions to the physical quantities cannot be neglected, and these contributions will give rise to deep conceptual changes. For example, ordinary special relativity, which is based on the Poincar\'e group, will no longer be true, and must be replaced by a new special relativity based on the de Sitter group. As a consequence, the ordinary notions of energy and momentum will change~\cite{aap}. The conserved momentum, for example, will now be obtained from the invariance of the physical system, not under translations, but under a combination of translations and proper conformal transformations. It will consequently be given by a combination of ordinary and proper conformal momenta. Energy, which is the time component of the momentum, will change accordingly.
Due to the fundamental role played by energy and momentum, these modifications will affect all branches of physics, including of course quantum mechanics. Although these effects may be negligible for small values of the cosmological term $\Lambda$, there are situations where this difference could become significant.

As an example of such situation, let us consider the following hypotheses. Taking into account that conformal symmetry has a relevant role at high energies, it is conceivable to assume that a high--energy phenomenon could modify the {\it local structure of space-time for a short period of time}, in such a way that the immediate neighborhood of a high energy phenomenon would depart from Minkowski and become a de Sitter --- or anti-de Sitter --- spacetime~\cite{mansouri}. According to this hypotheses, around a high--energy experiment there would exist a large $\Lambda$, and the local kinematics would consequently be ruled by the de Sitter special relativity. Concomitantly, the conformal symmetry would naturally acquire a relevant role. This scenario fits quite reasonably with the idea that a high--energy experiment should modify the local structure (texture) of spacetime. The important point is that the de Sitter special relativity gives a precise meaning to this change, opening up the door for a possible experimental verification~\cite{lorentz}. For an experiment with energy of the order of the Planck energy, the local value of $\Lambda$ would be of the order $\Lambda \sim 10^{66}~\mbox{cm}^{-2}$, which differs from the observed~\cite{obs} cosmological constant $\Lambda \sim 10^{-56}~\mbox{cm}^{-2}$ by roughly 120 orders of magnitude. The underlying spacetime in this case would approach a cone spacetime~\cite{confor}, which is transitive under proper conformal transformations only. In such extreme situation, the de Sitter special relativity would reduce to a conformal relativity, in which only the conformal notions of momentum and energy would survive. A very peculiar new quantum world would then emerge, whose physics has yet to be developed.

\begin{theacknowledgments}
The authors would like to thank FAPESP, CNPq and CAPES for 
financial support.
\end{theacknowledgments}



\begin{thebibliography}{30}

\bibitem{dssr}
R. Aldrovandi, J. P. Beltr\'an Almeida and J. G. Pereira, {\it Class. Quant. 
Grav.} {\bf 24}, 1385 (2007) ({\it Preprint} gr-qc/0606122).

\bibitem{guoatall}
H. Y. Guo, C. G. Huang, Z. Xu and B. Zhuo, {\it Phys. Lett.} A {\bf 331}, 1 (2204) 
({\it Preprint} hep-th/0403171).

\bibitem{lorentz}
R. Aldrovandi, J. P. Beltr\'an Almeida, C. S. O. Mayor and J. G. Pereira, {\it Lorentz Transformations in de Sitter Relativity} ({\it Preprint} gr-qc/0709.3947).

\bibitem{aap}
R. Aldrovandi, J. P. Beltr\'an Almeida and J. G. Pereira, {\it Int. J. Mod. Phys.} 
D {\bf 13}, 2241 (2004) ({\it Preprint} gr-qc/0405104).

\bibitem{confor}
R. Aldrovandi, J. P. Beltr\'an Almeida and J. G. Pereira, {\it J. Geom. Phys} {\bf 
56}, 1042 (2006)  ({\it Preprint} gr-qc/0403099).

\bibitem{carroll}
S. M. Carroll, {\it Living Rev.\ Rel.\ } {\bf 4}, 1 (2001) ({\it Preprint} astro-ph/0004075v2) 

\bibitem{mansouri}
F. Mansouri, {\it Phys. Lett.} B {\bf 538} 239 (2002) ({\it Preprint} hep-th/0203150)

\bibitem{weinberg}
S. Weinberg, {\it Gravitation and Cosmology}, Wiley, New York, 1972.

\bibitem{he}
S. W. Hawking and G. F. R Ellis, {\it The Large Scale Structure of
Space-Time}, Cambridge University Press, Cambridge, 1973.

\bibitem{livro}
R. Aldrovandi and J. G. Pereira, {\it An Introduction to
Geometrical Physics}, World Scientific, Singapore, 1995.

\bibitem{gursey}
F. G\"ursey, in {\it Group Theoretical Concepts and Methods in Elementary
Particle Physics}, ed. by F. G\"ursey, Istanbul Summer School of Theoretical
Physics, Gordon and Breach, New York, 1962.

\bibitem{inonu2}
E. In\"on\"u, in {\it Group Theoretical Concepts and Methods in Elementary
Particle Physics}, ed. by F. G\"ursey, Istanbul Summer School of Theoretical
Physics, Gordon and Breach, New York, 1962;
E. In\"on\"u and E. P. Wigner, {\it Proc. Natl. Acad. Scien.} {\bf 39}, 510 
(1953).

\bibitem{coleman}
S. Coleman, {\it Aspects of Symmetry}, Cambridge University Press, Cambridge, 
1985.

\bibitem{ugo}
S. Cacciatori, V.. Gorini, A. Kamenshchik and U. Moschela, {\it Conservation laws and scattering for de Sitter classical particles} ({\it Preprint} hep-th/0710.0315).

\bibitem{dix}
J. Dixmier, {\it Bulletin de la S. M. F.} {\bf 89}, 9 (1961) (available at http://www.numdam.org).

\bibitem{penrose}
R. Penrose, {\it Proc. Roy. Soc. (London)} {\bf 284 A}, 204 (1965).

\bibitem{deser}
S. Deser and A. Waldron, {\it Phys. Lett.} {\bf B 603}, 30 (2004) ({\it Preprint} hep-th/0408155).

\bibitem{ccj}
C. G. Callan, S. Coleman and R. Jackiw, {\it Ann. Phys. (NY)} {\bf 59}, 42 (1970).

\bibitem{massnomass}
V. Faraoni and F. I. Cooperstock, {\it Eur. J. Phys.} {\bf 19}, 419 (1998)
({\it Preprint} physics/9807056).

\bibitem{obs}
A. G. Riess {\it et al}, {\it Ap. J.} {\bf 116}, 1009 (1998);
S. Perlmutter {\it et al}, {\it Ap. J.} {\bf 517}, 565 (1999);
P. de Bernardis {\it et al}, {\it Nature} {\bf 404}, 955 (2000);
S. Hanany {\it et al}, {\it Ap. J. Letters} {\bf 545}, 5 (2000).
\end{thebibliography}
\end{document}